\documentclass[amsmath,amssymb,aps,prl,floatfix,twocolumn,superscriptaddress]{revtex4-2}

\usepackage{graphicx}
\usepackage{dcolumn}
\usepackage{bm}
\usepackage{physics}
\usepackage{booktabs}
\usepackage{amssymb}
\usepackage{amsthm}
\usepackage{multirow}
\usepackage[ruled,vlined]{algorithm2e}
\usepackage[table]{xcolor}
\usepackage{hyperref}

\theoremstyle{definition}
\newtheorem{definition}{Definition}

\newcommand{\mc}[1]{\mathcal{#1}}
\newcommand{\mr}[1]{\mathrm{#1}}

\newcommand{\qwith}{\quad \mathrm{with} \quad}

\newcommand{\iden}{\vb{1}}

\newcommand{\CZ}{\mathrm{CZ}}
\newcommand{\CZZ}{\mathrm{CZZ}}

\DeclareMathOperator*{\supp}{supp}

\begin{document}

\title{Fault-Tolerant Stabilizer Measurements in Surface Codes with Three-Qubit Gates}

\author{Josias Old}
\email{j.old@fz-juelich.de}
\affiliation{Institute for Theoretical Nanoelectronics (PGI-2), Forschungszentrum Jülich, Jülich, Germany}
\affiliation{Institute for Quantum Information, RWTH Aachen University, Aachen, Germany}

\author{Stephan Tasler}
\email{stephan.tasler@fau.de}
\affiliation{Physics Department, Friedrich-Alexander-Universität Erlangen Nürnberg, Germany}

\author{Michael J. Hartmann}
\affiliation{Physics Department, Friedrich-Alexander-Universität Erlangen Nürnberg, Germany}

\author{Markus Müller}
\affiliation{Institute for Theoretical Nanoelectronics (PGI-2), Forschungszentrum Jülich, Jülich, Germany}
\affiliation{Institute for Quantum Information, RWTH Aachen University, Aachen, Germany}

\date{\today}

\begin{abstract}
Quantum error correction (QEC) is considered a deciding component in enabling practical quantum computing. Stabilizer codes, and in particular topological surface codes, are promising candidates for implementing QEC by redundantly encoding quantum information.
While it is widely believed that a strictly fault-tolerant protocol can only be implemented using single- and two-qubit gates, 
several quantum computing platforms, based on trapped ions, neutral atoms and also superconducting qubits support native multi-qubit operations, e.g. using multi-ion entangling gates, Rydberg blockade or parallelized tunable couplers, respectively.
In this work, we show that stabilizer measurement circuits for unrotated surface codes can be fault-tolerant using single auxiliary qubits and three-qubit gates. 
These gates enable lower-depth circuits leading to fewer fault locations and potentially shorter QEC cycle times.
Concretely, we find that in an optimistic parameter regime where fidelities of three-qubit gates are the same as those of two-qubit gates, the logical error rate can be up to one order of magnitude lower and the threshold can be significantly higher, increasing from  $\approx 0.76 \%$ to  $\approx 1.05 \%$. 
Our results, which are applicable to a wide range of platforms, thereby motivate further investigation into multi-qubit gates as components for fault-tolerant QEC, as they can lead to substantial advantages in terms of time and physical qubit resources required to reach a target logical error rate.
\end{abstract}

\maketitle

\paragraph{Introduction.}--- 
Quantum error correction (QEC) is a crucial ingredient in enabling fault-tolerant quantum computation\,\cite{knill1998resilient, terhal2015quantum}. 
A widely used method involves \emph{stabilizer codes}, where repeatedly measuring a generating set of commuting \emph{stabilizer operators} reveals information on errors that occurred, forming the error-syndrome. 
While a stabilizer code with \emph{distance} $d$ can in principle correct for arbitrary weight-$t = \lfloor\frac{d-1}{2}\rfloor$ errors, implementing the stabilizer measurements with a noisy quantum circuit can reduce the number of correctable errors. Quantum gates or measurements, from which the error syndrome measurement circuits are formed, can be faulty, leading to a wrong diagnosis of the error. In addition, errors on auxiliary qubits can propagate to higher-weight errors on the data qubits (also called \emph{hook} errors), effectively reducing the distance.

Circuits where the distance of the underlying code is preserved, are deemed fault-tolerant (FT). Traditionally, as introduced e.g.~in Ref.\,\cite{aliferis2006quantum}, an error correction subroutine (or gadget) is called fault-tolerant if the \emph{weight} of the output error is bounded by the sum of the weight of the incoming error and the number of faults occurring during the protocol.
This turns (almost) any stabilizer readout with single, physical auxiliary qubits non-fault tolerant. There are numerous approaches to tackle these challenges, including \emph{Shor}, \emph{Steane} and \emph{Knill error correction}\,\cite{shor1996fault,steane1997active,knill1998resilient}. These methods rely on (fault-tolerantly) encoded auxiliary qubits to transversally read out the syndrome information using only single- and two-qubit gates. These approaches, as well as more recently developed flag-qubit based constructions\,\cite{yoder2017the, chao2018quantum, chamberland2018flag} make use of carefully crafted quantum circuits to prevent malicious error propagation.
There exist relaxed definitions of fault tolerance that include errors of weight $w>t$ that can still be corrected\,\cite{tansuwannont2022achieving}. Using these, it is known that \emph{any} stabilizer measurement circuit using two-qubit gates in the unrotated surface code is fault-tolerant\,\cite{manes2025distance}. The more qubit-efficient rotated surface codes, however, require a specific ordering of two-qubit gates to retain fault-tolerance\,\cite{tomita2014lowdistance}. 

To date, error correction cycles on rotated surface codes have been realized in trapped ions\,\cite{berthusen2024experiments} and superconducting architectures\,\cite{krinner2022realizing,zhao2022realization, acharya2023suppressing}. Recently, it has been shown that distance-$5$ and $7$ codes can be operated below the threshold, i.e. the logical error rate of the encoded logical qubit decreases with increasing code distance\,\cite{acharya2024quantum}.
Logical operations have been realized in error-detecting surface codes in superconducting architectures\,\cite{marques2022logical, zhang2024demonstrating}, with trapped ions~\,\cite{burton2024genons} and error-corrected entangling gates on neutral atom platforms\,\cite{bluvstein2024logical}.

These experimental QEC demonstrations use elementary gate sets consisting of single- and two-qubit gates. Most architectures, however, also allow for native implementation of multi-qubit gates, e.g.~using Rydberg blockade in neutral atoms\,\cite{lukin2001dipole,evered2023high}, multi-qubit Mølmer-Sørensen gates in trapped ions\,\cite{bermudez2017assessing} or optimized gate sequences for the control Hamiltonians of solid-state based platforms\,\cite{divincenzo2013multi,kim2022high} - according to common belief these would result in non-FT circuits. Recently, proposals have been made to use multi-qubit gates for surface code error correction in semiconductor spin qubits\,\cite{ustun2024single}, superconducting qubits\,\cite{reagor2022hardware,schwerdt2022comparing,tasler2025parralelized} and neutral atoms\,\cite{jandura2024surface}, While these references all use different noise models, they find that higher thresholds can be achieved in certain parameter regimes, but at the expense of losing strict fault-tolerance.

In this manuscript, we report on and explore the unexpected finding that stabilizer measurement circuits for unrotated surface codes using three-qubit gates can be fault-tolerant, even when considering three-qubit depolarizing noise channels on these gates.
We focus on the approach using single physical auxiliary qubits, repeatedly measuring stabilizers to gain confidence in measurement outcomes and carefully designing readout circuits to prevent malicious propagation of errors. 
We first recover standard results for surface-code syndrome readout circuits and then show how circuits using three-qubit $\CZZ$ gates are fault-tolerant, based on distinguishability of all  circuit faults up to order $t$ in the unrotated surface code.
We finally show comparative numerical studies that support the theoretical findings - the logical error rates of memory experiments scale $\propto p^{t+1}$ in the low physical error rate regime. 
We also find that, with optimistic assumptions on the noise strength, the parallel three-qubit $\CZZ$-gate based scheme outperforms sequential application of two-qubit $\CZ$ gates in terms of logical error rate. In particular, the threshold increases from $\approx 0.76 \%$ to $\approx 1.05 \%$.
These results can have practical significant impact for surface-code based QEC with ions, neutral atoms and solid-state platforms, lowering time and space resource requirements. 

\paragraph{Quantum error correction and surface codes.}--- 
In stabilizer error correction, the $+1$ eigenstates of a set of $n-k$ commuting Pauli operators span the $2^k$-dimensional logical subspace of an $n$-(physical) qubit Hilbert space\,\cite{gottesman1997thesis}. 
These operators generate the stabilizer group $\mc{S} = \{\langle S_i \rangle_{i=1}^{n-k}$, $\comm{S_i}{S_j} = 0 \,\, \forall i \neq j$, $\iden \notin \mc{S}\}$. 
Non-trivial Pauli operators on the codespace are all Pauli operators that commute with the stabilizers, but are not stabilizers themselves, i.e.~elements of the normalizer $\mc{N}_{\mc{P}}(\mc{S})$. 
The minimum weight of any such element is the code \emph{distance} $d$ and measures the performance of the code:  As stated earlier, a QEC code with distance $d$ can correct for arbitrary weight-$t$ with $t= \lfloor\frac{d-1}{2}\rfloor$ errors. The \emph{parameters} of a QEC codes are then the triple $[[n,k,d]]$. 

If a Pauli error $E \in \mc{P}^{\otimes n}$ occurs on the qubits, the outcomes of a projective measurement of a generating set of the stabilizer using auxiliary qubits yields its \emph{syndrome} $ \mathbf{s}(E) = (\langle S_i, E \rangle)_{i=1}^{n-k}$, which is decoded to correct for the error. Here, we denote by $\langle P, P' \rangle$ whether Pauli operators $P$ and $P'$ commute ($0$) or anticommute ($1$).
 
One of the most prominent families of stabilizer codes are topological surface codes\,\cite{dennis2002topological}. 
For a distance-$d$ surface code, data qubits are placed on the edges of a $d \times d$ square lattice. Stabilizer generators are defined on plaquettes fixing the $Z$-parity of qubits on adjacent edges and on vertices fixing the $X$-parity of qubits on emanating edges.
The resulting surface code has parameters ${[[d^2 + (d-1)^2, 1, d]]}$.
Additionally, one can cut the $(d-1)^2$ corner qubits to get the rotated surface code with parameters $[[d^2, 1, d]]$\,\cite{bombin2007optimal, horsman2012surface}, without reduction of the code distance.

\paragraph{Fault-tolerance of error correction circuits.}---
Loosely speaking,  a circuit implementing an error correction gadget using a QEC code with distance $d$ is fault-tolerant, if it takes at least $d$ distinct elementary (i.e.~$\mc{O}(p)$) faults to cause an undetected logical error. Formally, we use the following definition of circuit fault tolerance\,\cite{aliferis2006quantum,tansuwannont2022achieving}. 
We consider a circuit with (Clifford) gate locations, initializations and measurements. Additionally, we define \emph{detectors} as sums of measurements that are \emph{deterministic} in the absence of noise. Typically, a matrix $H \in \mathbb{F}_2^{n_{\mathrm{detectors}} \times n_{\mathrm{measurements}}}$ maps from bare measurement outcomes to detector flips. Every location $\ell \in \mc{L} = \{0,\dots,n_{\mr{locations}}-1\}$ of the circuit can be faulty, which is simulated by an ideal gate followed by a Pauli noise channel with error set $E(\ell)$.
For $n$-qubit gates and idling locations, we assume $n$-qubit depolarizing channels
\begin{align}
    \mc{E}_n(\rho) = (1-p) \rho + \frac{p}{4^n - 1} \sum_{i=1}^{4^n-1} P_n^i \rho P_n^i \\
    \qwith P_n^i \in \{\{I, X, Y, Z\}^{n} \setminus I^{\otimes n}\}. \nonumber
\end{align}
Single-qubit $Z$-basis initialization (measurement) is followed (preceded) by a bit-flip channel,
\begin{align}
    \mc{E}(\rho) = (1-p) \rho +p X \rho X.
\end{align}
We label each elementary fault with the location $\ell$ and the non-identity Pauli realization $P$ of the corresponding channel, s.t.~the set of all elementary faults is ${\mc{F}_1 = \{F_{\ell}^P\}_{\ell \in \mc{L}, P \in {E(\ell)}}}$. If the actual Pauli realization of the fault is not important, we typically drop the Pauli label.
Each fault is efficiently propagated through the circuit using Clifford simulation\,\cite{aaronson2004improved}, resulting in a (deterministic) \emph{error} $E(F)$, a vector of measurement outcomes $\mathbf{m}(F)$ and flipped detectors $\mathbf{D}(F) = H \mathbf{m}(F)$.
An order-$w$ \emph{fault path} $\mathbf{F}$ is a set of $w$ faults at distinct locations, $\mathbf{F} = \{F_\ell\}_{\ell \in L \subseteq \mc{L}, \abs{L} = w}$.
The errors and flipped detectors that result from all faults in a fault path can be calculated as $E(\mathbf{F}) = \prod_{F \in \mathbf{F}} E(F)$ and $\mathbf{D}(\mathbf{F}) = \bigoplus_{F \in \mathbf{F}} \mathbf{D}(F)$, where $\oplus$ denotes element-wise summation modulo $2$.

If any fault path up to order $t$ in $p$ leads only to correctable errors (and could hence be corrected by a subsequent ideal round of stabilizer measurement), then we say the circuit is \emph{distance preserving} and fault tolerant. This is captured by the notion of \emph{distinguishable fault sets}\,\cite{tansuwannont2022achieving}.
\begin{definition}
    We call the collection of all possible fault paths of order $\leq w$ the \emph{fault set} $\mc{F}^{(w)}$. A fault set is \emph{distinguishable} if for any pair of fault paths $\mathbf{F}_i, \mathbf{F}_j \in \mc{F}^{(w)}$, either
    \begin{enumerate}
        \item $\mathbf{s}(E(\mathbf{F}_i)) \neq \mathbf{s}(E(\mathbf{F}_j))$,
        \item $E(\mathbf{F}_i) \sim_{\mathcal{S}} E(\mathbf{F}_j)$, or
        \item $\mathbf{D}(\mathbf{F}_i) \neq \mathbf{D}(\mathbf{F}_j)$.
    \end{enumerate}
    That is, any two fault paths either result in errors with different syndromes (1.), in stabilizer-equivalent errors (2.), or in different detector flips during the protocol (3.).  
\end{definition}

\begin{figure}
    \centering
    \includegraphics[width=\linewidth]{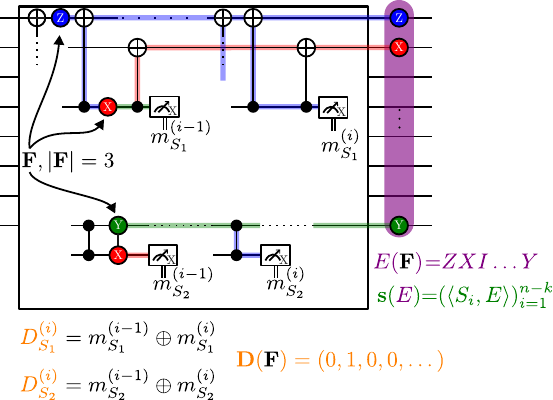}
    \caption{Objects used to determine distinguishability of a fault set. The box represents some (Clifford) circuit with measurements. It is encoded in an error-correcting code with stabilizer generators $\{S_i\}_{i=1}^{n-k}$. During the circuit, sets of deterministic measurements specify the detectors $\{D\}$. For an error-correction gadget, these are typically the parities of consecutive stabilizer measurements. The exemplarily shown fault path $\mathbf{F}$ of weight $w=3$ can be efficiently propagated using Clifford simulation. This results in a vector of flipped detectors $\mathbf{D}$ and a final error $E$. The ideal syndrome of $E$ is $s(E) = (\langle S_i , E \rangle)_{i=1}^{n-k}$. }
    \label{fig:distinguishability}
\end{figure}

We visualize the quantities used in this approach in Fig.\,\ref{fig:distinguishability}.  
The key technical observation (Prop.\,1 of Ref.\,\cite{tansuwannont2022achieving}) is that iff  $\mc{F}^{(w)}$ is distinguishable, then the smallest fault path leading to a direct, undetected logical error is in $\mc{F}^{(2w + 1)}$.
This implies that a circuit employing a code of distance $d$ with elementary fault set $\mc{F}_1$ is fault-tolerant iff the fault set $\mc{F}^{(t)}$ is distinguishable.
This is a similar situation compared to fault tolerance in detector error models, see e.g. Ref.\,\cite{derks2024designing}.

\begin{figure}
    \centering
    \includegraphics[width=0.8\linewidth]{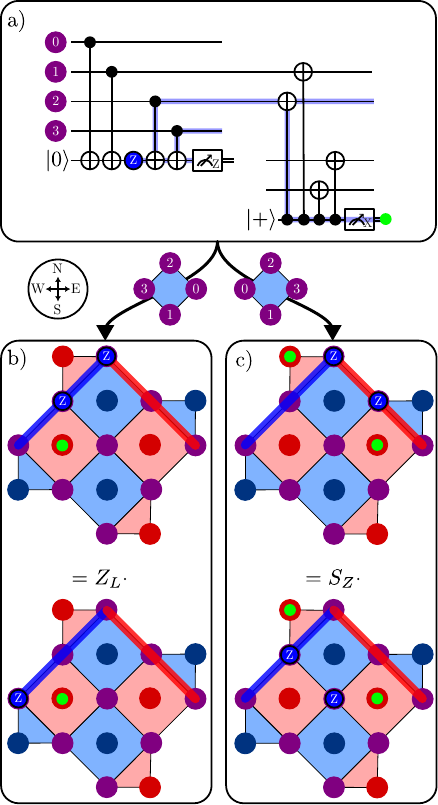}
    \caption{Detail of stabilizer measurement circuits for the distance-$3$ rotated surface code. a) A $Z$ fault on the ancilla of a $Z$-stabilizer measurement can propagate to two data qubits, indicated by the blue highlighting. The propagated fault is detected in the next round of $X$-stabilizer measurements. b) and c) show the effect for two different orderings of the gates. In the rotated surface codes, we draw the $X$-($Z$-)logical operators as thick red (blue) lines and draw flipped detectors with a light green dot. b) If the last two gates act on the north (N) and west (W) qubit, the final $Z$ error is parallel to the $Z$-logical operator. There is another first order fault with the same syndrome that is logically inequivalent implying a non-distinguishable fault set $\mc{F}^{(1)}$. c) For an ordering towards the  north (N) and east (E) qubit, however, faults with the same flipped detectors are stabilizer-equivalent. The fault set $\mc{F}^{(1)}$ is thus distinguishable.}
    \label{fig:rotated_cz_dist}
\end{figure}

\paragraph{Distinguishable Fault Sets of Surface Code Syndrome Measurement Circuits.}---
We first recall stabilizer measurement circuit constructions for rotated and unrotated surface codes using single- and two-qubit gates. 
In Fig.\,\ref{fig:rotated_cz_dist} a), we show a circuit to measure a $Z$-stabilizer of the distance-$3$ rotated surface code. The order of $\CZ$s determines whether the fault set $\mc{F}^{(1)}$ is distinguishable or not. If a fault on the auxiliary qubit (\emph{hook error}) propagates to an error parallel to the corresponding logical operator, there exists another fault path with the same syndrome and, crucially, with the same detector outcomes. 
In Fig.\,\ref{fig:rotated_cz_dist} b), we show that the errors resulting from the pair of faults are also not stabilizer-equivalent, such that they are indistinguishable.
When propagating to qubits orthogonal to the logical operator (Fig.\,\ref{fig:rotated_cz_dist} c)), no other fault path in $\mc{F}^{(1)}$ has the same syndrome or non-stabilizer-equivalent error, ensuring distinguishability. 

\begin{figure}
    \centering
    \includegraphics[width=0.8\linewidth]{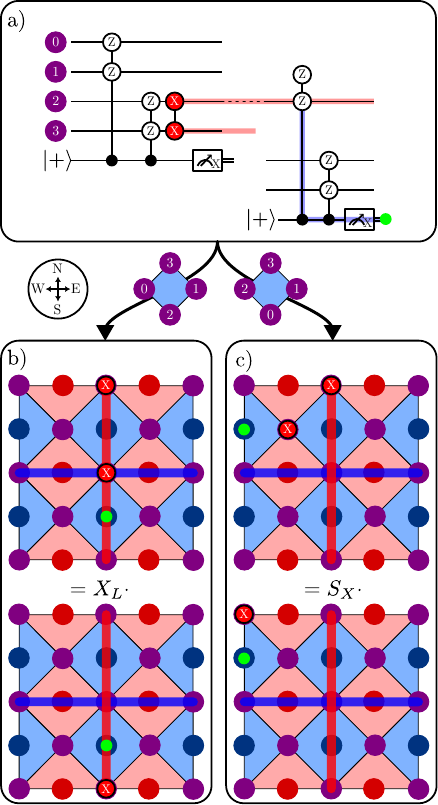}
    \caption{Detail of stabilizer measurement circuits for the distance $3$ unrotated surface code, implemented with three-qubit $\CZZ$ gates as described in the main text. a) For a uniform depolarizing noise model, there also exist elementary two-qubit $X$-faults on the support of $Z$-stabilizer generators - contrary to the two-qubit gate situation. These faults are captured by a subsequent $Z$-stabilizer measurement. b) and c) show the effect for two different orderings of the gates. In the unrotated surface codes, we draw the $X$-($Z$-)logical operators as thick red (blue) lines and draw flipped detectors with a light green dot. b) If the last three-qubit gates act on the horizontal (WE) and vertical (NS) qubits, the final $X$ error is parallel to the $X$-logical. There is another first-order fault with the same syndrome, which is logically inequivalent implying a non-distinguishable fault set $\mc{F}^{(1)}$. c) For a diagonal ordering towards the south-east (SE) and north-west (NW) qubits, however, faults of the same order with the same flipped detectors are stabilizer-equivalent. The fault set $\mc{F}^{(1)}$ is distinguishable.}
    \label{fig:unrotated_czz_dist}
\end{figure}

For the unrotated surface code, it was noted in Ref.\,\cite{dennis2002topological} that hook errors are not as damaging as naively expected. This can be explained by realizing that unrotated surface codes can be constructed as hypergraph product codes of two (classical) repetition codes\,\cite{tillich2013quantum}.
Ref.\,\cite{manes2025distance} showed that for any hypergraph product code, any order of two-qubit entangling gates gives a distance preserving stabilizer measurement circuit. 
This is equivalent to $\mc{F}^{(t)}$ being distinguishable.
The fundamental reason for that is that every stabilizer generator (of Pauli $P$) overlaps with any minimum-weight logical operator (of Pauli $P$) on at most a single qubit, as can be seen in Fig.\,\ref{fig:unrotated_czz_dist}\,b). 
Because during the measurement of a Pauli $P$-type stabilizer generator, only Pauli-$P$ errors are propagated on the data qubits and any propagated error has at most $1$ overlap with a (minimum weight) logical operator, $d$ faults are required to make a direct logical error. By linearity, $t = \frac{d-1}{2}$ faults are distinguishable.

Now, we extend these arguments for circuits using three-qubit $\CZZ$ gates, as shown in Fig.\,\ref{fig:unrotated_czz_dist} a) for a $Z$-stabilizer measurement. 
To manage hook errors, two data qubits on the support of the three-qubit gate have to reflect the ordering in the two-qubit case, i.e. rotated surface codes require propagation orthogonal to the logical operators and unrotated surface codes are robust against hook errors. 

In addition to hook errors, there are now also elementary faults of arbitrary two-qubit Pauli operators on data qubits involved in the $\CZZ$ gate.
These potentially turn $\mc{F}^{(t)}$ indistinguishable. In rotated surface codes, there is no way to retain distinguishability: If the direction of the $\CZZ$ is chosen such that e.g. Pauli-$Z$ hook errors propagate orthogonally to the $Z$-logical operator, the elementary $XX$ fault on the data qubits of the $\CZZ$ gate is parallel to the $X$-logical operator and vice versa.

In the unrotated surface code, we differentiate between horizontal/vertical (West-East, WE or North-South, NS) and diagonal (North-East, NE or North-West, NW) qubits of a stabilizer that are supported on the $\CZZ$ gate, see Fig.\,\ref{fig:unrotated_czz_dist} b) and c).
The horizontal and vertical qubits overlap with $Z$- and $X$-logical operators on two positions. Therefore only $\frac{d+1}{2}$ faults are needed to make a direct undetected logical error. Equivalently, there exist two indistinguishable fault paths in order $\mc{F}^{(t)}$, also shown in Fig.\,\ref{fig:unrotated_czz_dist} b).

Choosing the diagonal data qubits and adding the corresponding faults to $\mc{F}^{(t)}$, however, does not render it indistinguishable.
The reason for that is that all these newly introduced faults have either a new syndrome (because they are the union of two distinguishable single-qubit faults) or are stabilizer-equivalent. We show this for an example in Fig.\,\ref{fig:unrotated_czz_dist} c). Again, the argument is that such faults overlap with minimum-weight logical generators on only a single position. Contrary to rotated surface codes, this also holds for $X$-faults on the support of $Z$-stabilizers and vice versa.
All minimum-weight logical generators of the unrotated surface codes have support only on the 'outer' $d \times d$ square lattice. Whenever a multi-qubit gate involves one data qubit of the outer and one from the inner  $d-1 \times d-1$ lattice, there are still $d$ of these faults required for an undetected logical error and the fault set $\mc{F}^{(t)}$ remains distinguishable.

We numerically verify the distinguishability of fault sets for circuits implemented with $\CZZ$s using Alg.\,\ref{alg:check_ft} (cf. Supplemental Material), where we first simulate the effect of each elementary fault, and then construct all up-to-order-$t$ combinations of faults until we find indistinguishable fault paths. We summarize the results of our exhaustive checks in Tab.\,\ref{tab:dist_checks} of the Supplemental Material. 
These numerical results complement the above arguments and confirm that, while rotated surface codes have their distance reduced to $d_{\mathrm{eff}} = \frac{d-1}{2}$, the NE and NW three-qubit gate stabilizer measurement circuits for unrotated surface codes are fault-tolerant with the full effective distance $d_{\mathrm{eff}} = d$.

\paragraph{Memory Experiments.}---
We now investigate the error-correction performance of the three-qubit gate protocol. To that end, we perform full circuit level noise memory experiments using $\texttt{stim}$\,\cite{gidney2021stim}. 
We construct circuits for rotated and unrotated~surface codes similar to the circuits in the experimental Ref.\,\cite{krinner2022realizing}, i.e.~we implement $X$- and $Z$-stabilizer readout circuits one after another, with measurement of $X$-($Z$-)ancillary qubits during the $Z$-($X$-)entangling gate cycle. In the $\CZ$-protocol, we order the gates orthogonal to the logical operators, i.e. South--East--West--North (SEWN) for $X$- and South--West--East--North (SWEN) for $Z$-type stabilizers. When using the three-qubit $\CZZ$-gate, we order both $X$- and $Z$-stabilizers NW-SE, which we further discuss in the Supplemental Material.
We consider a noise model similar to the SI1000 model of Ref.\,\cite{gidney2022benchmarking} (SI) shown in Tab.\,\ref{tab:noise_strengths}. For simplicity, and without restriction of generality, we also set idling noise to $0$ (NI), to focus on the bare effect of $\CZZ$ gates followed by three-qubit depolarizing channels. We take the same noise strength for $\CZZ$ gates as for the $\CZ$ gates which rests on the assumption that noise is dominated by the gate duration and a parallelized $\CZZ$ gate takes the same time as a single $\CZ$ gate. This can be regarded as an optimistic, though not unrealistic parameter regime and the associated QEC performances represents loosely speaking an upper limit of the potential offered by the three-qubit based approach. We discuss the QEC performance for complementary parameter regimes of higher three-qubit gate error rates in the Supplemental Material.
In Ref.\,\cite{tasler2025parralelized}, we show how a three-qubit $\CZZ$ gate with equal gate times as a two-qubit $\CZ$ gate can be realized with transmon-based qubits in superconducting circuits.  
For further details on the implementation of the circuits, the multi-qubit gates and error channels, refer to the Supplemental Material, where we also elaborate on the choice of decoder. In the following, we use pymatching (pm)\,\cite{higgott2021pymatching} for NI simulations and beliefmatching (bm)\,\cite{higgott2023improveddecoding} for SI.

\begin{table}
    \centering
    \caption{Noise strengths of circuit operations used for the simulations. We use a superconducting inspired noise model (SI) and a noise model without idling noise (NI). }
    \label{tab:noise_strengths}
    \begin{tabular}{ccc}
    \toprule
      & \multicolumn{2}{c}{noise strength}   \\
     \multirow{-2}{*}{gate} & SI & NI   \\
    \midrule
        $\CZ, \CZZ$  & $p$  & $p$  \\
        $\mr{H}$ & $p/10$ & $p/10$  \\ 
        $\mr{Init}_Z$ & $2p$ & $2p$  \\
        $\mr{M}_Z$ & $5p$ & $5p$  \\ 
        $\mr{Idle}, \mr{Idle}_{\mr{M,Init}}$ & $p/10$ & $0$  \\
    \bottomrule
    \end{tabular}
\end{table}

\begin{figure*}
    \centering
    \includegraphics[width=0.95\linewidth]{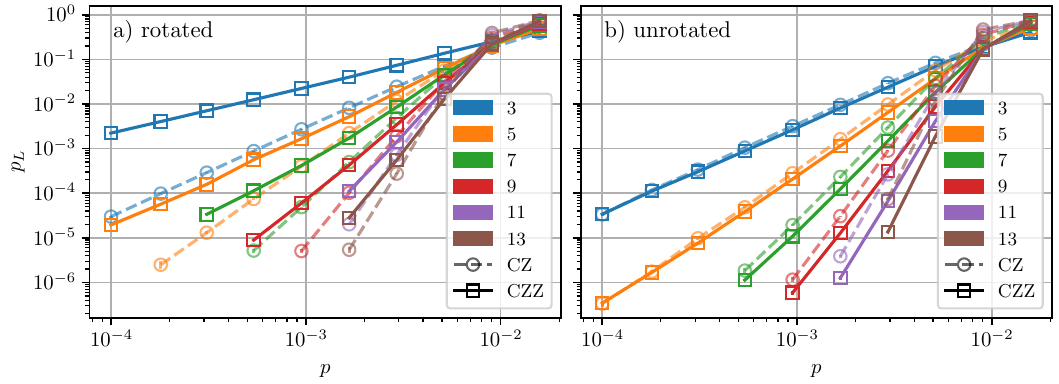}
    \caption{Logical error rates for implementations using $\CZ$ (circle, dashed) and $\CZZ$ gates (square, solid) in the NI model. The color coding represents codes with increasing distance. a) In a \emph{rotated} surface code, using three qubit $\CZZ$ gates breaks fault tolerance which can be seen by the different scaling for small physical error rates $p$. Additionally, the logical error rate is higher, e.g. about an order of magnitude for distance-$9$ codes at $p \approx 10^{-3}$. b) In \emph{unrotated} surface codes, however, the logical error rate barely changes when replacing $\CZ$s with $\CZZ$s (using the same error strength, but $3$- instead of two $2$- qubit depolarizing channels). For low $p$, we also observe a flattening of the scaling. This, however, is due to the suboptimal decomposition of elementary faults for decoding with pymatching, as explained in the main text. Error bars are standard Monte Carlo errors and can be smaller than the symbol used.}
    \label{fig:plot_use_czz_sweep_noidle}
\end{figure*}

The results for the NI model are shown in Fig.\,\ref{fig:plot_use_czz_sweep_noidle} for the rotated (left) and unrotated (right) surface codes.
The logical error rates of \emph{rotated} surface codes using $\CZZ$ show a scaling $\propto p^{\frac{t+1}{2}}$ for small physical error rates, consistent with the expected effective circuit fault distance $d_{\mathrm{eff}} = \frac{d-1}{2}$. These circuits are therefore not fault-tolerant in the strict sense defined above. The threshold value, however, is increased compared to the  circuits implementing the stabilizer measurement using conventional $\CZ$ gates. 
This is consistent with the observations of Refs.\,\cite{reagor2022hardware,jandura2024surface} and is attributed to the smaller number of fault locations. In particular, the threshold is about $1.21-1.38$ times larger. This is close to the ratio of number fault locations in $\CZ$ and $\CZZ$ circuits of $1.25$. We discuss this in more detail in the Supplemental Material. 
The logical error rates for the \emph{unrotated} surface codes are comparable in magnitude and show the expected scaling $\propto p^{t+1}$ through most of the simulated range of physical errors. 
At low physical error rates $p < 10^{-3}$, we can see a flattening of the logical error rate for the $\CZZ$-based protocol. A closer investigation of faults that lead to logical failures in that regime reveals that this is merely related to a suboptimal decomposition of faults into matchable faults for decoding - and not to fault-tolerant breaking processes. We show an example of such a failed decomposition in the Supplemental Material.

\begin{figure*}
    \centering
    \includegraphics[width=0.95\linewidth]{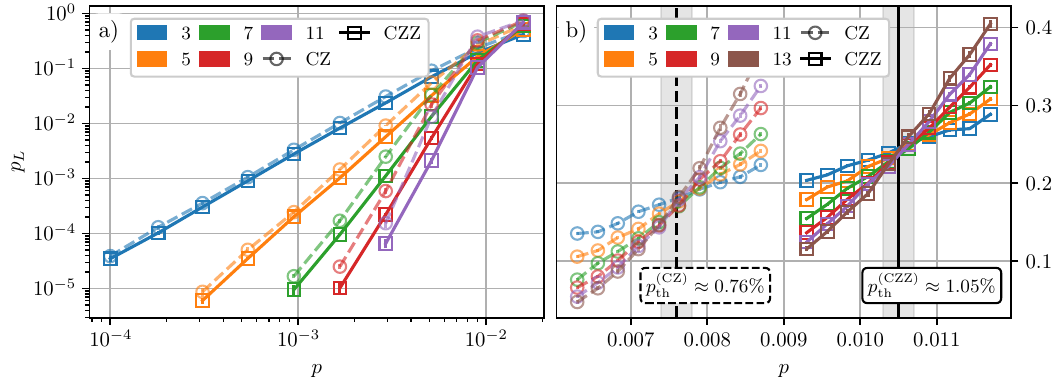}
    \caption{Logical error rates for implementations using $\CZ$ (circle, dashed) and $\CZZ$ gates (square, solid). We assume idling noise strength of $p/10$. The color coding represents codes with increasing distance. a) Distance-$3$ and $5$ codes correct for all faults up to order $t$, resulting in a scaling of the logical error rate $p_L \propto p^{t+1}$ as expected for a fault-tolerant protocol. Due to fewer idling locations using three-qubit gates, the logical error rate is up to 50\% lower for physical error rates in the range of $10^{-2}$ to $10^{-3}$.  b) The threshold is increased from $p^{(\mathrm{CZ})}_{\mathrm{th}} \approx 0.76 \%$ to  $p^{(\mathrm{CZZ})}_{\mathrm{th}} \approx 1.05 \%$. Decoded using beliefmatching with $d$ iterations of BP before matching. Error bars are standard Monte Carlo errors and can be smaller than the symbol used. Thresholds are obtained using the finite-size scaling ansatz described in the Supplemental Material.}
    \label{fig:use_czz_thresholdsweep}
\end{figure*}

For the SI model, i.e.~introducing idling noise with strength $p/10$, the advantage of more compact circuits becomes even more apparent, as shown in Fig.\,\ref{fig:use_czz_thresholdsweep}. Additionally, we optimize the beliefmatching decoder for distance $3$ and $5$, to handle the remaining non-corrected faults.
Therefore, our simulations confirm the fault-tolerance of both the $\CZ$- and $\CZZ$-gate based implementation. 
Notably, with these optimistic noise parameters, the threshold using $\CZZ$ gates is much higher compared to the implementation with $\CZ$ gates and is increased from  $p^{(\mathrm{CZ})}_{\mathrm{th}} \approx 0.76 \pm 0.02 \%$ to  $p^{(\mathrm{CZZ})}_{\mathrm{th}} \approx 1.05 \pm 0.02 \%$. This is a $38\%$ increase that we again attribute to the smaller number of fault locations. An overview of all obtained thresholds is shown in Tab.\,\ref{tab:thresholds}.

\begin{figure}
    \centering
    \includegraphics[width=0.95\linewidth]{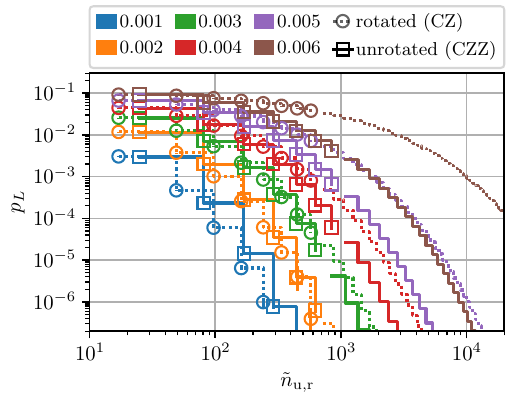}
    \caption{Number of physical qubits required to reach a target logical error rates for $d$ rounds of stabilizer measurements under circuit level noise (NI) of strengths $p \in (0.1\%-0.6\%)$. Comparing fault-tolerant rotated surface code circuits using $\CZ$ gates (circle, dotted) and fault-tolerant unrotated surface codes with  $\CZZ$ gates (square, solid). Extrapolated by the exponential fit function described in the main text. From $p > 0.3\%$, the unrotated surface codes consistently require fewer physical qubits compared to rotated surface codes in order to reach a target logical error rate.}
    \label{fig:plot_p_L_n_p_lattice_style_SI1000}
\end{figure}

\paragraph{Qubit-resource comparison for FT QEC.}---
Typically, for practical implementations, the rotated surface code is preferred over its unrotated counterpart because of the smaller qubit count.  
The unrotated surface code uses $\tilde{n}_{\mathrm{u}} = 4d^2-4d+1$ physical qubits (data + auxiliary) compared to the rotated surface code with $\tilde{n}_{\mathrm{r}} = 2d^2-1$, essentially halving the amount of physical qubits required for large distances. This often results in better performance of rotated surface codes for a fixed number of physical qubits. In the following, we show that in a realistic noise regime, the unrotated surface code with $\CZZ$ gates can actually achieve a target logical error rate with \emph{fewer} physical qubits than the traditional rotated surface code circuits using $\CZ$s.
We plot the logical error rate achieved by surface codes against the number of physical qubits in Fig.\,\ref{fig:plot_p_L_n_p_lattice_style_SI1000} and compare a $\CZ$-rotated with a $\CZZ$-unrotated implementation for near-term experimentally relevant sub-threshold physical error rates in the regime of $0.1-0.7\%$. 
Our key observation is that there is a threshold physical error rate above which the unrotated surface codes require \emph{less} physical qubits to achieve a target logical error rate. 
We find this error rate at $p^{(\circ)}_{\mathrm{th}} \approx 0.3\%$.
To extrapolate to small logical error rates, we fit 
\begin{align}
    p_L(n) = c_0 \qty(\frac{p}{c_1})^{c_2 \sqrt{n}}
\end{align}
and find fit parameters shown in Tab.\,\ref{tab:fit_params_n_qubits}.
At $p = 0.2 \%$ and to reach a target logical error rate of $10^{-6}$, considered a value where quantum computers can solve useful tasks like factoring of numbers\,\cite{preskill1998reliable}, a rotated surface code has to have distance $d_{\mathrm{r}} = 17$, resulting in a total of $\tilde{n}_{\mathrm{r}} = 557$ physical qubits, slightly lower than the $\tilde{n}_{\mathrm{u}} = 625$ qubits in unrotated surface codes, corresponding to a distance $d_{\mathrm{u}}=13$. 
At $p = 0.3 \%$, however, rotated surface codes require a total of $\tilde{n}_{\mathrm{r}} = 1457\,(d_{\mathrm{r}} = 27)$ physical qubits, about $34 \%$ more than unrotated surface codes with  ${\tilde{n}_{\mathrm{u}} = 1089}\,(d_{\mathrm{u}} = 17)$.

\begin{table}
   \centering
   \caption{Thresholds of rotated and unrotated surface codes. We compare $\CZ$ or $\CZZ$ circuits for plots shown in the main text, without idling noise and decoded using pymatching (pm), with idling noise and decoded with beliefmatching (bm). The uncertainty is extracted from a finite size scaling analysis, detailed in the Supplemental Material. }
   \label{tab:thresholds}
   \begin{tabular}{ccccc}
      \toprule
                                  &    & $p_{\mr{th}}^{(\CZ)}$ & $p_{\mr{th}}^{(\CZZ)}$ & $p_{\mr{th}}^{(\CZZ)}/p_{\mr{th}}^{(\CZ)}$ \\
      \midrule
                                  & pm & $0.7 \pm 0.02\%$& $0.85 \pm 0.01\%$& $1.21$\\
      \multirow{-2}{*}{rotated}   & bm & $0.75 \pm 0.02\%$& $1.00 \pm 0.02\%$& $1.33$                                             \\
      \midrule
                                  & pm & $0.71 \pm 0.02\%$& $0.932 \pm 0.008\%$& $1.31$\\
      \multirow{-2}{*}{unrotated} & bm & $0.76 \pm 0.02 \%$& $1.05 \pm 0.02\%$& $1.38$                                             \\
      \bottomrule
   \end{tabular}
\end{table}

\paragraph{Conclusion.}---
We have investigated the application of three-qubit gates for stabilizer measurement circuits in rotated and unrotated surface codes. We have shown how $\CZZ$-gate based circuits in unrotated surface codes are fault-tolerant, even for an adversarial uniform depolarizing noise model on the support of the three-qubit gates.  We have observed that in an optimistic parameter regime for the achievable fidelity of the multi-qubit gates, logical error rates are lower and thresholds of these circuits are higher, due to a smaller number of fault locations. In particular, we have shown that for near-term experimentally relevant error rates, unrotated surface codes with $\CZZ$ gates can be the less-qubit intensive version to reach a target logical error rate.

Our results with uniform depolarizing noise also suggest that a closer investigation of the noise channels of multi-qubit gates can lead to further improvements. In a parallel work, we construct a $\CZZ$ gate that effectively realizes two parallel $\CZ$ gates in transmon-based superconducting circuits\,\cite{tasler2025parralelized}. We show that an optimization for minimizing fault-tolerance breaking faults can reduce the logical error rates also for rotated surface codes. 

These results put multi-qubit gates back on the map as potentially highly valuable building blocks for QEC which are compatible with FT circuit design principles. In particular, investigating which other QLDPC codes like lifted product or bivariate bicycle codes\,\cite{xu2024constant,bravyi2024high, old2024lift} allow for fault-tolerant parallelized stabilizer readout is an interesting open question and might improve practical feasibility of such larger classes of QEC codes.

\paragraph{Acknowledgements.}---
We gratefully acknowledge support by the European Union’s Horizon Europe research and innovation programme under Grant Agreement No.~101114305 (“MILLENION-SGA1” EU Project). This research is also part of the Munich Quantum Valley (K-8), which is supported by the Bavarian state government with funds from the Hightech Agenda Bayern Plus. We additionally acknowledge support by the BMBF project MUNIQC-ATOMS (Grant No. 13N16070) and the BMBF project GeQCoS (Grant No. 13N15684).
MM also acknowledges support for the research by the European ERC Starting Grant QNets through Grant No. 804247, and by IARPA and the Army Research Office, under the Entangled Logical Qubits program through Cooperative Agreement Number W911NF-23-2-0216. The views and conclusions contained in this document are those of the authors and should not be interpreted as representing the official policies, either expressed or implied, of IARPA, the Army Research Office, or the U.S. Government. The U.S. Government is authorized to reproduce and distribute reprints for Government purposes notwithstanding any copyright notation herein.
MM and JO acknowledge support by the Deutsche Forschungsgemeinschaft (DFG, German Research Foundation) under Germany’s Excellence Strategy “Cluster of Excellence Matter and Light for Quantum Computing (ML4Q) EXC 2004/1” 390534769. 
The authors gratefully acknowledge the computing time provided to them at the NHR Center NHR4CES at RWTH Aachen University (Project No. p0020074). This is funded by the Federal Ministry of Education and Research and the state governments participating on the basis of the resolutions of the GWK for national high performance computing at universities.

\bibliography{bibliography.bib}

\clearpage

\onecolumngrid
\appendix

\begin{center}
\textbf{\large Supplemental Material}
\end{center}

\setcounter{equation}{0}
\setcounter{figure}{0}
\setcounter{table}{0}
\setcounter{page}{1}
\makeatletter
\renewcommand{\theequation}{S\arabic{equation}}
\renewcommand{\thefigure}{S\arabic{figure}}
\renewcommand{\thetable}{S\arabic{table}}

\section{Memory experiments} 
In a  $Z$-($X$-) memory experiment for CSS stabilizer codes, we initialize into the $+1$ eigenstate of $Z$-($X$-) logical operators, keep it alive and measure it as follows:
\begin{enumerate}
    \item initialize data qubits in product state $\ket{0}^{\otimes n}$ ($\ket{+}^{\otimes n}$),
    \item using auxiliary qubits, measure ($X$- and $Z$-) stabilizers $n_{\mr{rounds}}$ times,
    \item destructively measure all data qubits in the $Z$-($X$-) basis, yielding measurement results $\{m_i\}_{i=0}^{n-1}$.
\end{enumerate}
Details in the circuit implementation can lead to differences in the logical error rates of $Z$- and $X$-memory experiments. We therefore always simulate both bases and report the overall logical error probability 
\begin{align}
    p_L = 1 - (1-p_L^{(X)})(1-p_L^{(Z)}) \label{eqn:logical_error_rate}
\end{align}
assuming independence of $X$- and $Z$- logical error rates.

\section{Stabilizer measurement circuits}
There are numerous ways to schedule and order the entangling gates required for the projective measurement of the stabilizer generators. 
In unrotated surface codes, the order does not influence the fault-distance of the circuits as described in the main text and shown below. In rotated surface codes, an implementation with two-qubit entangling gates requires a scheduling such that the last two gates interact with data qubits that are \emph{orthogonal} to the logical operator corresponding to the Pauli type of the measured stabilizer\,\cite{tomita2014lowdistance}. We therefore also expect an ordering dependence of the logical error rate for a protocol using three-qubit $\CZZ$ gates.

\begin{figure*}
    \centering
    \includegraphics[width=\linewidth]{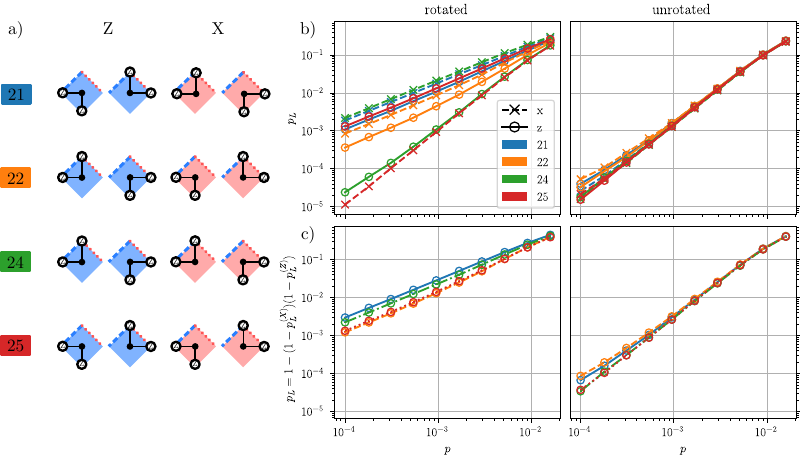}
    \caption{a) Orderings $21,22,24$ and $25$ of three-qubit CZZ gates for $Z$- and $X$-stabilizer measurements. On the plaquettes, we draw in dashed (dotted) lines the overlap of a $Z$- ($X$-) logical operator in the rotated surface code. b) Logical error rate of $X$- and $Z$-basis memory experiments on distance $d=7$ rotated surface codes. For orderings 21 and 22,  these show a symmetric behaviour. Orderings 24 and 25 perform differently for the two bases as described in the text. c) Combining the $X$- and $Z$- logical error rates shows that the logical error rates are very similar across the orderings. }
    \label{fig:cz_directions}
\end{figure*}

We show the different schedules in Fig.\,\ref{fig:cz_directions} a), and the respective logical $X$- and $Z$- error rates for a distance-$3$ surface code implemented with the CZZ gates using these schedule in Fig.\,\ref{fig:cz_directions} b). We use a uniform depolarizing noise model as described in the main text. 

These results confirm the ordering-independence of the unrotated surface code circuits. For the rotated surface codes and an ordering that is orthogonal (21) or parallel (22) to the respective logical, the logical error rates are also symmetric in $X$ and $Z$. 
We can, however, turn one of the memory experiments fault-tolerant by ordering both, $Z$- and $X$- stabilizer measurements orthogonal to a fixed logical operators. For ordering $24$, e.g., all gates are orthogonal to the $X$-logical. 
This allows for a distance-preserving protection against $X$-logical errors, such that the $Z$-basis memory experiments shows the expected FT scaling (here $\propto p^4$ for $d=7$). This holds analogously for ordering $25$ with Pauli operators $X$ and $Z$ interchanged. In Fig.\,\ref{fig:cz_directions} c), we show the combined logical error rate (Eqn.\,\ref{eqn:logical_error_rate}). There is still a small difference in orderings. 
For the rotated surface codes, we attribute this to finite size effects as in some orderings the boundary stabilizers are measured using $\CZ$ gates and two time steps, whereas in others, they are measured using $\CZZ$ gates and one timestep. In the unrotated surface codes, a closer investigation reveals that certain faults that trigger more than $2$ detectors are decomposed by \texttt{stim} in such a way that the fault-distance is reduced. We show and example of such a fault in Fig.\,\ref{fig:failing_decomposition}. 

\begin{figure*}
    \centering
    \includegraphics[width=\linewidth]{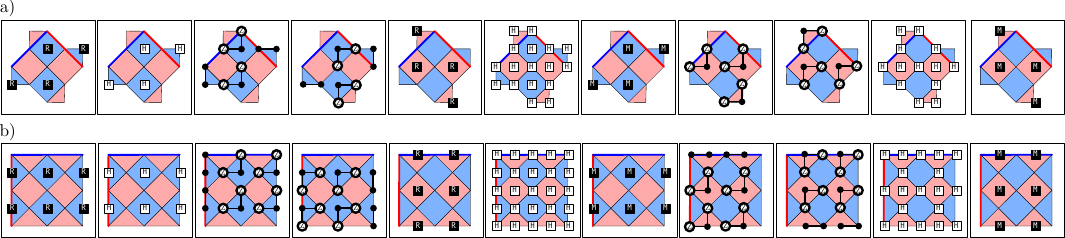}
    \caption{One round of stabilizer measurements of distance-$3$ surface codes in the a) rotated and b) unrotated implementation. $X$- and $Z$-stabilizers are drawn in light red and blue, respectively. We also indicate minimum-weight logical operators. $\CZZ$ gates always have their control qubit on the ancillary qubit in the center of a plaquette. The implementation with $\CZZ$ gates requires two timesteps of entangling gates per Pauli type, compared to $4$ timesteps for an equivalent implementation with $\CZ$ gates. }
    \label{fig:circuit_stab_readout}
\end{figure*}

We therefore simulate the circuits using ordering $24$, shown in Fig.\,\ref{fig:circuit_stab_readout} for a single measurement of the stabilizers of distance $d=3$ rotated and unrotated surface codes. Note that we restrict the gate set to $Z$-basis initialization, measurement and entangling gates, as well has single-qubit Hadamard gates for basis changes. This is inspired by native gate sets in superconducting qubit architectures\,\cite{krinner2022realizing, acharya2023suppressing}.

\section{Verification of fault tolerance}
We verify fault tolerance in the sense of Ref.\,\cite{tansuwannont2022achieving} numerically using Alg.\,\ref{alg:check_ft}. The results of exhaustive checks are summarized in Tab.\,\ref{tab:dist_checks} and confirm the fault-tolerance of three-qubit gates circuits on unrotated surface codes. Additionally, we checked the effective distance of the memory experiment circuits for unrotated surface codes up to distance $d=9$ in stim using \texttt{circuit.search\_for\_undetectable\_logical\_errors(\dots)}.

\begin{algorithm}
\DontPrintSemicolon
\KwIn{Quantum Circuit $\mc{C}$, Elementary Fault set $\mc{F}_1$, Parity Check Matrix $H \in \mathbb{F}_2^{m \times n}$, Logical Matrix $L \in \mathbb{F}_2^{k \times n}$, Detector Matrix $D\in \mathbb{F}_2^{n_D \times n_m}$, $t$}
\KwOut{Distinguishability of fault set $\mc{F}^{(t)}$}

\For{$F \in \mc{F}_1$}{
$E(F), \mathbf{m}(F) = \mathrm{Propagate}(F)$ \;
$\mathbf{d}(F) = D \mathbf{m}(F)$\;
}
construct all up to order $t$ combinations of faults $\mc{F}^{(t)}$\;
initialize graph
    $\mathfrak{G} = (\mathfrak{V} = \{\mc{D} = \{\},\mc{F}^{(t)},\mc{E} = \{\},\mc{S} = \{\}\}, $\,
    $\phantom{\,\mathfrak{G} = (}\mathfrak{E} = \{\}) $\;
\For{$\mathbf{F} \in \mc{F}^{(t)}$ }{
  $\mc{D} \leftarrow \mathbf{d}(\mathbf{F})$ (cumulative detector flips)\;
  $\mc{E} \leftarrow E(\mathbf{F})$ (final Pauli error)\;
  $\mc{S} \leftarrow s = HE$ (final syndrome) \;
  $\mathfrak{E} \leftarrow \{(\mathbf{d}(\mathbf{F}), \mathbf{F}), (\mathbf{F}, E(\mathbf{F})), (E(\mathbf{F}), s)\}$
} 
\For{$\mathbf{d} \in \mc{D}$}{
 Divide all $E$ neighbors of $\mathbf{F}$ neighbors of $\mathbf{d}$ in stabilizer equivalent subsets\;
 \If{$\exists$ more than one subset}{
 \If{any neighboring syndrome $s$ is connected to more than one subset.}    {\KwRet{False}}
 }
}
\KwRet{True}
\caption{Distinguishability of fault sets of a quantum circuit} \label{alg:check_ft}
\end{algorithm}

\begin{table}
    \centering
    \caption{Largest distinguishable fault sets of different surface code configurations and $\CZZ$ readout directions/schedules. Exhaustively verified using Alg.\,\ref{alg:check_ft}.}
    \label{tab:dist_checks}
    \begin{tabular}{cccc}
       \toprule
       lattice  & $d$ & $\CZZ$ direction & largest distinguishable fault set  \\
       \midrule
       \multirow{6}{*}{rotated}  & \multirow{3}{*}{$3$} & NE & \cellcolor{red!25} $\{\}$  \\
         &  & NW & \cellcolor{red!25} $\{\}$  \\
         &  & NS & \cellcolor{red!25} $\{\}$  \\
         \cmidrule{2-4}
         & \multirow{3}{*}{$5$} & NE & \cellcolor{orange!25} $\mc{F}^{(1)}$  \\
         &  & NW &\cellcolor{orange!25} $\mc{F}^{(1)}$  \\
         &  & NS &\cellcolor{orange!25} $\mc{F}^{(1)}$  \\
         \midrule
       \multirow{6}{*}{unrotated}  & \multirow{3}{*}{$3$} & NE & \cellcolor{green!25}$\mc{F}^{(1)}$  \\
         &  & NW & \cellcolor{green!25}$\mc{F}^{(1)}$  \\
         & & NS & \cellcolor{red!25}$\{\}$  \\
         \cmidrule{2-4}
         & \multirow{3}{*}{$5$} & NE & \cellcolor{green!25}$\mc{F}^{(2)}$  \\
         &  & NW & \cellcolor{green!25}$\mc{F}^{(2)}$  \\
         &  & NS & \cellcolor{orange!25}$\mc{F}^{(1)}$ \\
       \bottomrule 
    \end{tabular}
\end{table}

\section{Detector error model}

For CSS stabilizer codes, denote by $S_{p,i}^t$ the outcome of measurement of the Pauli $p$-type stabilizer $i$ at round $t$. We declare detectors in the 'standard' way, i.e. 
\begin{align}
    D_{z,i}^{t} = S_{z,i}^{t} \oplus S_{z,i}^{t-1} &\qfor t \in \{0,\dots ,n_{\mr{rounds}}\},\\
    &\phantom{\qfor}i \in \{0, n_{Z\text{-}\mr{stabilizers}}-1\} \\
    &\qwith S_{z,i}^{-1} = 0 \\
    &\qand S_{z,i}^{n_{\mr{rounds}}} = \bigoplus_{q \in \supp{S_{z,i}}} m_q,\\
    D_{x,i}^{t} = S_{x,i}^{t} \oplus S_{x,i}^{t-1} &\qfor t \in \{1,\dots ,n_{\mr{rounds}}-1\} ,\\
    &\phantom{\qfor}i \in \{0, n_{X\text{-}\mr{stabilizers}}-1\}.
\end{align}
In words, subsequent measurements of stabilizers should give the same result and are used as detectors.
Additionally, if we initialize and measure in the $Z$-basis, we compare the first $Z$- stabilizer measurement to the initialized value ($0$) and the last $Z$- stabilizer measurement to the stabilizer eigenvalues reconstructed from the final single-qubit measurements. For $X$-type memory experiments, swap the $Z$ and $X$-labels.

Next, we annotate the relevant ($Z$- or $X$-) observables, 
\begin{align}
    L^{(i)} = \bigoplus_{q \in \supp{L^{(i)}_{z/x}}} m_q.
\end{align}

Finally, we define a noise model as a list of possible elementary faults in order to be able to generate a \emph{detector error model}.
This detector error model consists of
\begin{itemize}
    \item a \emph{detector matrix} $D \in \mathbb{F}_2^{n_{\mr{detectors}} \times n_{\mr{elementary\:faults}}}$ with $D_{ij} = 1 \iff$ elementary fault $j$ triggers detector $i$, 
    \item a \emph{logical matrix} $L \in \mathbb{F}_2^{n_{\mr{observables}} \times n_{\mr{elementary\:faults}}}$ with $L_{ij} = 1 \iff$ elementary fault $j$ flips observable $i$ and
    \item a \emph{prior vector} $\vec{p} \in \mathbb{R}^{n_{\mr{elementary\:faults}}}$ where $p_i$ is the probability of elementary fault $i$ occurring.  
\end{itemize}
The detector error model can be generated by placing each elementary fault in the circuit and evaluating its action on detectors and logicals/observables.

Benchmarking the code then amounts to
\begin{enumerate}
    \item Sample elementary faults $\vec{f}$ according to the probability distribution defined by $\vec{p}$,
    \item record the 
    \begin{enumerate}
    \item syndrome $\vec{s} = D \vec{f}$,
    \item observation $\vec{l} = L \vec{f}$,
    \end{enumerate}
    \item decode the syndrome to get a fault-guess $\vec{f}' = \mathrm{Dec}(\vec{s}, D, \vec{p})$ such that $D \vec{f}' = \vec{s}$,
    \item calculate the prediction $\vec{l}' = L \vec{f}'$,
    \item record a logical error iff $\vec{l}' \neq \vec{l}$.
\end{enumerate}
Note that the choice of decoder depends not only on the code, but also on the noise model as we will see below.

\section{Representing $n$-qubit depolarizing error in \texttt{stim}}
There is no native way to include depolarizing channels acting on $n > 2$ qubits in \texttt{stim}.
However, \texttt{CORRELATED\_ERROR}s can be any Pauli error acting on any number of qubits. Despite its name, a correlated error can act as an independent error if it is not followed by an \texttt{ELSE\_CORRELATED\_ERROR}.
Consider as an example a single-qubit uniform depolarizing channel with, such that the total probability of an $X$-error is $p$. If simulated by three independent channels applying $X,Y$ and $Z$ with probability $p_{\mathrm{ind}}$, it has to hold that $p_{\mathrm{ind}} (1-p_{\mathrm{ind}})^2 + p_{\mathrm{ind}}^2 (1-p_{\mathrm{ind}}) = p$, because $YZ \propto X$. This solves to $p_{\mathrm{ind}} = \frac{1}{2} - \frac{1}{2}\left(1-4p\right)^{\frac{1}{2}}$ and can be generalized to
\begin{align}
p_{\mr{ind}} = \frac{1}{2} - \frac{1}{2}\left(1 - \frac{4^n}{4^n-1}p\right)^{2^{1-2n}},
\end{align}
where $n$ is the number of qubits the depolarizing channels acts on\,\cite{gidney2020decorrelated}. In order to simulate a three-qubit depolarizing channel of strength $p$, we therefore reproduce its probability distribution using independent errors with this rescaled error probability.
This results in errors of the form
\begin{align}
    \texttt{CORRELATED\_ERROR(p\_ind) X1*X2*X3} \nonumber \\
    \texttt{CORRELATED\_ERROR(p\_ind) X1*X2*Y3} \nonumber \\
    \texttt{CORRELATED\_ERROR(p\_ind) X1*X2*Z3} \nonumber \\
    \texttt{\dots} \nonumber
\end{align}

\section{Decomposing the detector error model for a matching-based decoder} \label{sss:decoding_simplified_noise_model}
A matching-based decoder only works with elementary faults flipping at most $2$ detectors. 
Consider the set of elementary faults $F$, and assume, some flip more than $2$ detectors. Then, if there is a (potentially smaller) set of (different) elementary faults $F'$ that can generate any fault in $F$, then any possible syndrome generated by a circuit under noise model defined by $F$ is in principle reachable by a circuit with noise model defined by $F'$. We can then use the detector matrix $D'$ to decode. In particular, if we find a basis of elementary faults that only flip $\leq 2$ detectors, we can build a matching graph based on $D'$. This, however, comes at a price of suboptimal decoding because correlations might get lost.

The canonical example for this procedure is a depolarizing error model in the surface code. While elementary faults are $X,Y$ and $Z$ with the same probability, only $X$- and $Z$- faults trigger two detectors and $Y$- faults trigger $4$. 
To be able to use a matching based decoder, one can therefore use the fact that $XZ\propto Y$ to generate the detector error model for decoding using a simplified noise model only consisting of $X$- and $Z$-faults.

The benchmarking procedure using $D$ and $D'$ is
\begin{enumerate}
    \item generate two detector error models $(D, L, \vec{p})$ for noise model $F$ and  $(D', L', \vec{p}')$ for the simplified noise model $F'$, 
    \item sample elementary faults $\vec{f}$ according to the probability distribution defined by $\vec{p}$,
    \item record the 
    \begin{enumerate}
    \item syndrome $\vec{s} = D \vec{f}$,
    \item observation $\vec{l} = L \vec{f}$,
    \end{enumerate}
    \item decode the syndrome using the simplified detector error model to get a fault-guess $\vec{f}' = \mathrm{Dec}(\vec{s}, D', \vec{p}')$ such that $D' \vec{f}' = \vec{s}$,
    \item calculate the prediction using the simplified logical matrix $\vec{l}' = L' \vec{f}'$,
    \item record a logical error if $\vec{l}' \neq \vec{l}$.
\end{enumerate}

\begin{figure*}
    \centering
    \includegraphics[width=\linewidth]{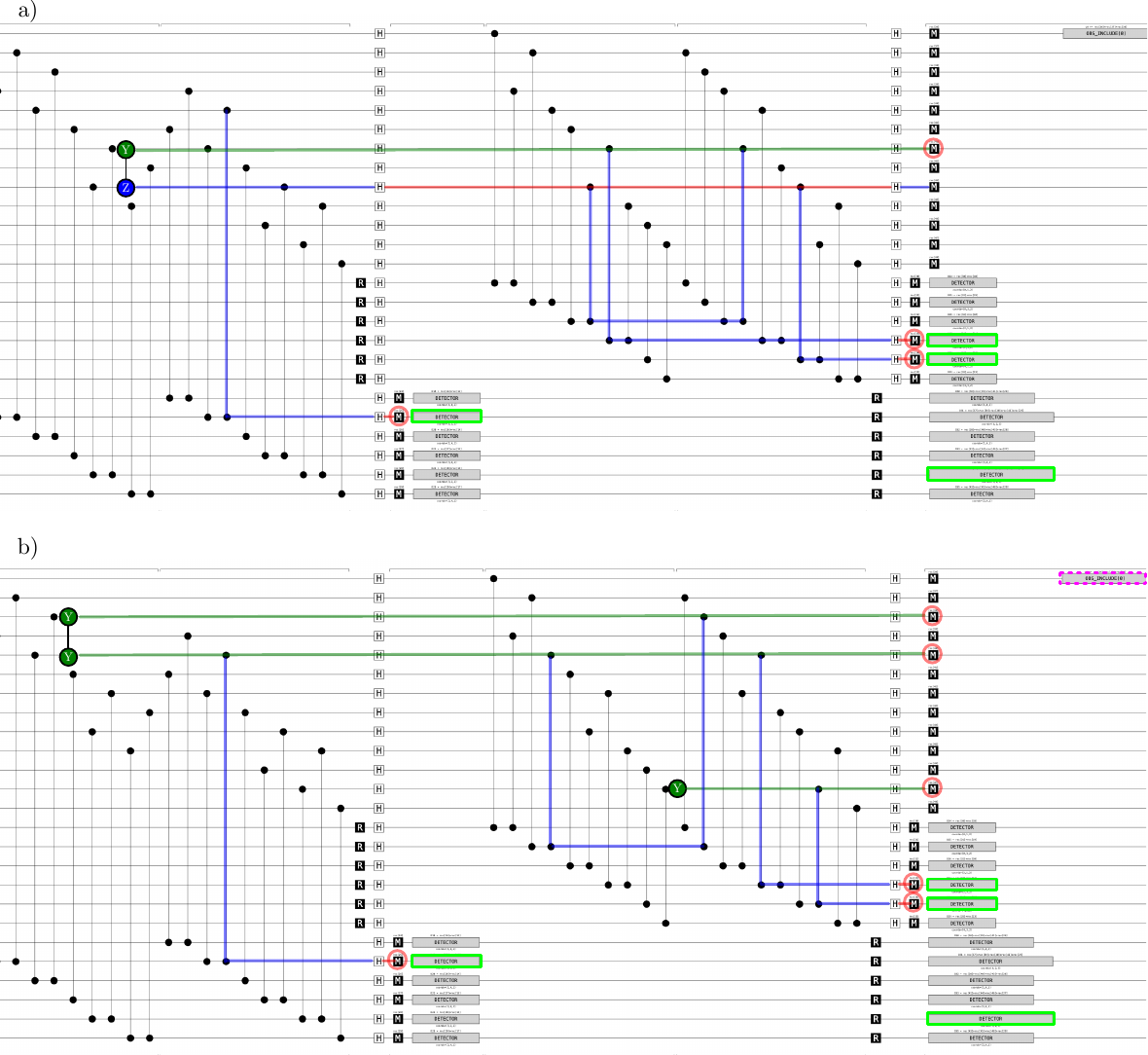}
    \caption{Exemplary failing decomposition of faults elementary faults in stabilizer measurement circuit of the a distance-$3$ unrotated surface code. We draw the Pauli operator propagation with red, green and blue highlights for $X$, $Y$ and $Z$, respectively. A flipped measurement is indicated by a red circle, a flipped detector by a light green rectangle. The flipped observable is marked with a dashed pink rectangle. a) An elementary $YZ$ fault flipping $4$ detectors and no observable. b) The decomposed detector error model fed to the matching decoder suggests two elementary faults ($YY$ and $Y$), that flip the same detectors. However, this correction also flips the observable, leading to a failure. }
    \label{fig:failing_decomposition}
\end{figure*}

With three-qubit depolarizing errors, several elementary faults flip more then two detectors. 
We use \texttt{stim}'s built-in detector error model generation that heuristically decomposes faults to turn the model matchable. 
While for surface code circuits with two-qubit $\CZ$ gates this procedure is distance-preserving, some decompositions prevent pymatching from correcting all up to order $p$ faults. We show one example  of such a decomposition in Fig.\,\ref{fig:failing_decomposition}. We find that using beliefmatching\,\cite{higgott2023improveddecoding}, where the weight of edges of the full decoding graph are updated using $d$ iterations of belief propagation before decomposing, can correct for all elementary faults in the distance $3$ and $5$ codes.

\section{Fit parameters}
We model the logical error rate for a fixed physical error rate $p$  extrapolate to small logical error rates, we fit 
\begin{align}
    p_L(n) = c_0 \qty(\frac{p}{c_1})^{c_2 \sqrt{n}}
\end{align}
and find fit parameters shown in Tab.\,\ref{tab:fit_params_n_qubits}.

\begin{table}[t]
   \centering
   \caption{Fit parameters for $p_L(n)$, see main text.}
   \label{tab:fit_params_n_qubits}
   \begin{tabular}{lccccc}
      \toprule
      $p$                      & Surface Code & $c_0$  & $c_1$ & $c_2$ \\
      \midrule
      \multirow{2}{*}{$0.001$} & rotated      & 0.08 &  0.009  &    0.33 \\
                               & unrotated    & 0.14 &  0.0066 &    0.38 \\
      \midrule
      \multirow{2}{*}{$0.002$} & rotated      & 0.11 &  0.0056 &    0.46 \\
                               & unrotated    & 0.2  &  0.0074 &    0.39 \\
      \midrule
      \multirow{2}{*}{$0.003$} & rotated      & 0.12 &  0.0056 &    0.51 \\
                               & unrotated    & 0.2  &  0.0087 &    0.35 \\
      \midrule
      \multirow{2}{*}{$0.004$} & rotated      & 0.12 &  0.0063 &    0.45 \\
                               & unrotated    & 0.2  &  0.0075 &    0.43 \\
      \midrule
      \multirow{2}{*}{$0.005$} & rotated      & 0.12 &  0.0063 &    0.51 \\
                               & unrotated    & 0.2  &  0.0077 &    0.45 \\
      \midrule
      \multirow{2}{*}{$0.006$} & rotated      & 0.12 &  0.0068 &    0.37 \\
                               & unrotated    & 0.19 &  0.0076 &    0.57 \\
      \bottomrule
   \end{tabular}
\end{table}

\section{Number of fault locations}
We estimate the increase in threshold by comparing the number of fault locations of the circuits.
For all of our memory experiment circuits, there are $n$ faulty initializations, $n$ faulty final measurements, $2dn$ faulty single-qubit Hadamard gates for data qubit basis changes. 
A weight-$w$ stabilizer measurement (without the basis changes) has $w+4$ fault locations using $\CZ$ gates. For $\CZZ$-gates, this reduces to $\lceil w/2 \rceil +4$ fault locations.
Distance-$d$ rotated surface codes have $(d-1)^2$ weight-$4$ bulk and  $2(d-1)$ weight-$2$ boundary stabilizers.
Distance-$d$ unrotated surface codes have $2(d-1)(d-2)$ weight-$4$ bulk and  $4(d-1)$ weight-$3$ boundary stabilizers.
This yields the number of fault locations summarized in Tab.\,\ref{tab:n_fault_locations} and their (asymptotic) ratios
\begin{align}
    \frac{n_{\text{fl}}^{\text{rotated, CZ}} }{n_{\text{fl}}^{\text{rotated, CZZ}} } &\xrightarrow{d \to \infty} \frac{5}{4} \\
    \frac{n_{\text{fl}}^{\text{unrotated, CZ}} }{n_{\text{fl}}^{\text{unrotated, CZZ}} } &\xrightarrow{d \to \infty} \frac{5}{4}.
\end{align}

\begin{table}
    \setlength{\tabcolsep}{10pt}
    \centering
    \caption{Number of fault-locations ($n_{\text{fl}}$) of memory-experiments  for rotated and unrotated surface codes of distance $d$. }
    \label{tab:n_fault_locations}
    \begin{tabular}{llr}
        \toprule
        surface code & gate & number of fault locations $n_{\text{fl}}$ \\
        \midrule
     & $\CZ$  & $10d^3-2d^2-4d$ \\
        \multirow{-2}{*}{Rotated}    & $\CZZ$ & $8d^3-4d$ \\
        \midrule
          & $\CZ$  & $20d^3-20d^2+2d+2$ \\
        \multirow{-2}{*}{Unrotated}  & $\CZZ$ & $16d^3-12d^2-2d+2$ \\
        \bottomrule
    \end{tabular}
\end{table}

\section{Fidelity of CZZ gate}
\begin{figure*}
    \centering
    \includegraphics[width=\linewidth]{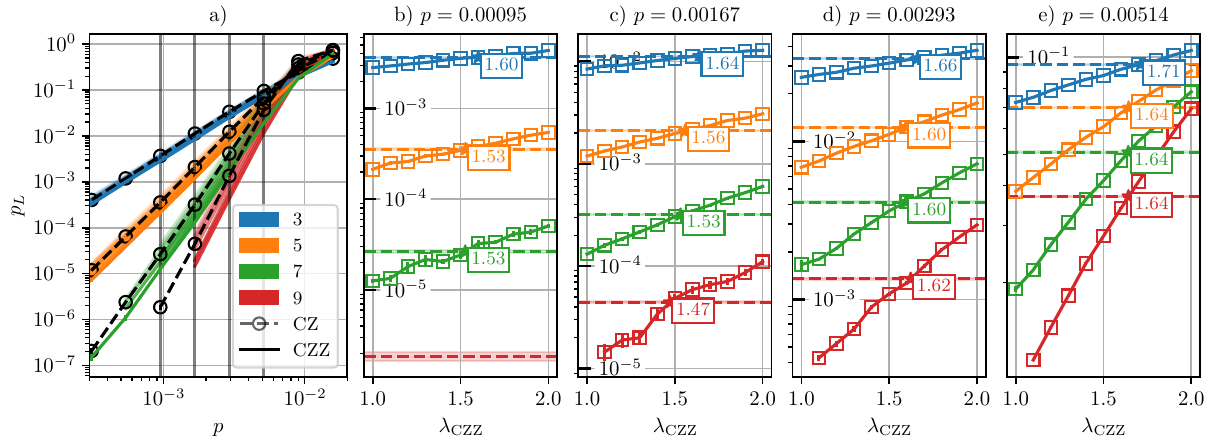}
    \caption{Influence of $\lambda_{\mathrm{CZZ}} = \frac{p_{\mathrm{CZZ}}}{p_{\mathrm{CZ}}}$ on the logical error rate. Unrotated surface codes with circuit level noise and an idling strength of $p_{\mathrm{idle}} = 0.1 p$, decoded using pymatching. a) In black, we show logical error rate of an implementation with two-qubit $\CZ$ gates. The colored range are logical error rates for implementations with three-qubit $\CZZ$ gates for $\lambda_{\mathrm{CZZ}} \in [1.0,2.0]$. b) - e) shows cuts along different physical error rates. In these, we show the intersection of the logical error rates of $\CZZ$ circuits (square marker) with $\CZ$ circuits (dashed horizontal line). In general, a smaller distance and a larger physical error rate allow for larger $\lambda_{\mathrm{CZZ}}$, i.e. worse-performing three-qubit gates. For $\lambda_{\mathrm{CZZ}} < 1.5$, the multi-qubit circuits consistently outperform the two-qubit gate based circuits. }
    \label{fig:plot_lCZZ}
\end{figure*}

In the main text, we assume that the three-qubit gates have the same failure rate as two-qubit gates $p_{\mathrm{CZZ}} = p_{\mathrm{CZ}} = p$.
In current state-of-the-art experiments, typical three-qubit gate fidelities are, however, worse than two-qubit fidelities. Ref.\,\cite{evered2023high} achieves fidelities of $F_{\mathrm{CCZ}} = 97.9(2)\%$ compared to $F_{\mathrm{CZ}} = 99.52(4)\%$ for neutral atoms. In superconducting qubits, an iToffoli gate with fidelity $F_{\mathrm{iCCX}} = 98.26(2)\%$ has been demonstrated\,\cite{kim2022high}.
In the following, we investigate how much larger the error rate of the $\CZZ$ gate can be compared to the $\CZ$ gate, to still provide an advantageous QEC performance, by scaling $p_{\mathrm{CZZ}} = \lambda_{\mathrm{CZZ}} p$. In
Fig.\,\ref{fig:plot_lCZZ}, we show results by comparing different $\lambda_{\mathrm{CZZ}} \in [1.0,2.0]$ for the unrotated surface codes with the  circuit level noise model and an idling noise strength of $p_{\mathrm{idle}} = 0.1 p$. We find that for $\lambda_{\mathrm{CZZ}} \leq 1.5$, the logical error rate using the multi-qubit gate is still lower than when using two-qubit gates, across all error rates.

\clearpage
\section{Finite size scaling analyses of thresholds}
\begin{figure*}[b]
    \centering
    \includegraphics[width=\linewidth]{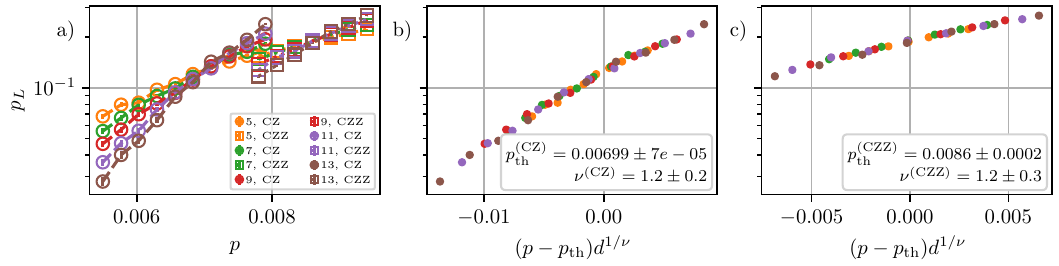}
    \includegraphics[width=\linewidth]{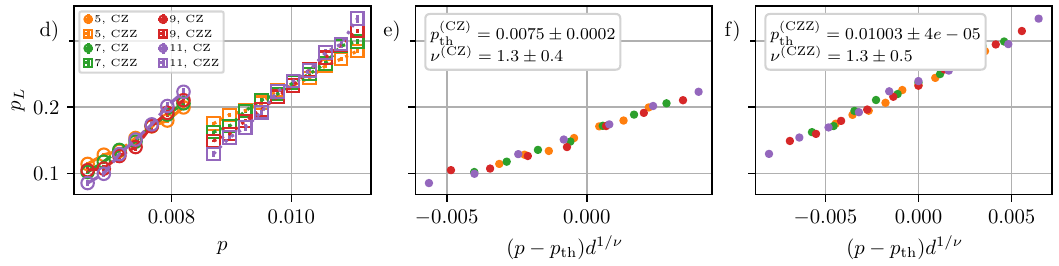}
    \includegraphics[width=\linewidth]{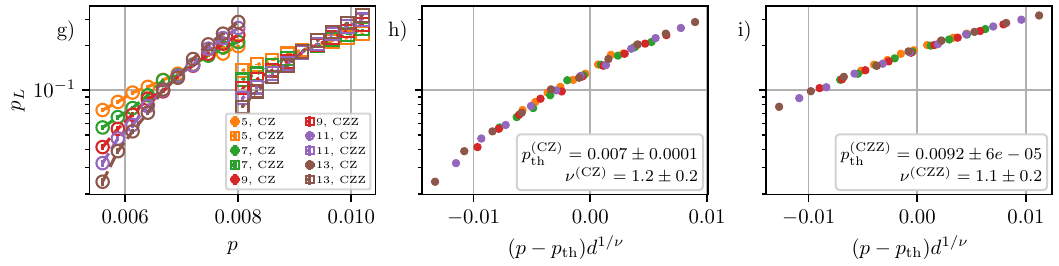}
    \includegraphics[width=\linewidth]{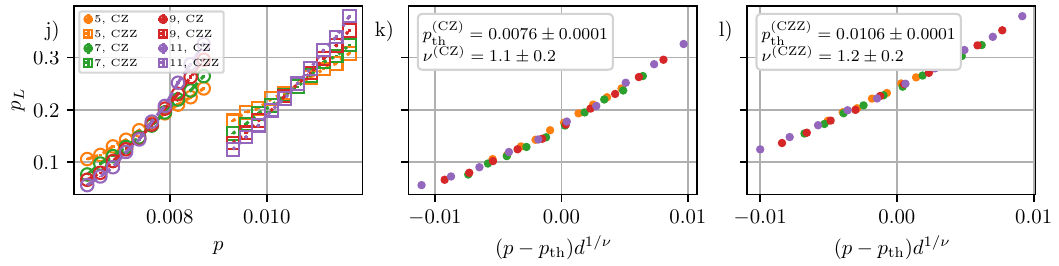}
    \caption{Threshold plots and data collapse optimized with \texttt{pyfssa}. a)-c) Rotated surface codes, no idling noise,  pymatching. d)-f) Rotated surface codes, idling noise,  beliefmatching. g)-i) Unrotated surface codes, no idling noise,  pymatching. j)-l) Unrotated surface codes, idling noise,  beliefmatching.}
    \label{fig:plot_thr_fss}
\end{figure*}

We perform finite-size scaling analysis using the \texttt{pyfssa} package\,\cite{melchert2009autoscale, sorge2015pyfssa}. We make the ansatz for a scaling in the form of $p_L(p) = f((p - p_{\mathrm{th}})d^{1/\nu})$, where $d$ is the distance and $f(x)$ a linear dimensionless function. The \texttt{pyfssa} package then determines the critical exponents and thresholds that give the best data collapse. We show the results in Fig.\,\ref{fig:plot_thr_fss}.

\end{document}